\documentclass[11pt]{article}
\newcommand{\bq}{\begin{equation}}
\newcommand{\eq}{\end{equation}}
\newcommand{\ba}{\begin{eqnarray}}
\newcommand{\ea}{\end{eqnarray}}

\newcommand{\nl }{ \nonumber  }

\newcommand{\p}{\partial}

\newcommand{\h}{\hspace{.5cm}}
\newcommand{\s}{\sigma}

\begin{document}
%\titlepage
\begin{flushright}
{\bf IC/2001/149}
\end{flushright}
\begin{center}
\vspace*{1cm} {\bf EXACT STRING SOLUTIONS AND THE OPEN STRING - D-BRANE
SYSTEM IN NON-CONSTANT BACKGROUND FIELDS I. \vspace*{.5cm}\\ P. Bozhilov
\footnote{E-mail: bozhilov@ictp.trieste.it, p.bozhilov@shu-bg.net,
bojilov@thsun1.jinr.ru}},
\\ {\it Department of Theoretical Physics,
\\ "K. Preslavsky" University, 9712 Shoumen, Bulgaria}
\end{center}
\vspace*{.5cm}

%====abstract===========
New exact string solutions in non-constant background fields are found and
it is shown that some of them are compatible with the boundary conditions
for the open string - D-brane system. Extension of the constraint algebra
is proposed and discussed.\\

PACS number(s): 11.25.-w, 11.27.+d, 11.25.Sq, 11.10.Lm

%====end of abstract====

\vspace*{.5cm}

%%%%%%%%%%%%%%%%%%%%%%%%%%%%%%%%%%%%%%%%%%%%%%%%%%%%%%%%%%%%%%%%%%%%%%%%%%%%%
\section{Introduction}
%%%%%%%%%%%%%%%%%%%%%%%%%%%%%%%%%%%%%%%%%%%%%%%%%%%%%%%%%%%%%%%%%%%%%%%%%%%%%
\hspace{1cm} Obtaining exact solutions of the nonlinear probe string
equations of motion and constraints in variable external fields is by all
means an interesting task with many possible applications. One such
application is connected with the recent investigations of the open string
- D-brane system in non-constant background fields
\cite{CHK99}-\cite{HN01}. Of course, in this case, one is forced to use
different approximations in order to get explicit results. That is why, it
is interesting to see to what extent our knowledge about the existing {\it
exact} string solutions can help us in considering this dynamical system.
More concretely, which kind of solutions are compatible with the mixed
boundary conditions characterizing the open string - D-brane system.

In this letter, we consider a few types of exact solutions of the probe
string equations of motion and constraints in non-constant background metric
and NS-NS two-form gauge field. Then we check their compatibility with the
open string - D-brane boundary conditions. It turns out that there exist
one type of exact string solutions, which gives also nontrivial solution of
the mixed boundary conditions for the open string - D-brane system.
After that, we reinterpret the conditions for existence of such solutions as
a set of constraints and compute their Poisson bracket algebra.

%%%%%%%%%%%%%%%%%%%%%%%%%%%%%%%%%%%%%%%%%%%%%%%%%%%%%%%%%%%%%%%%%%%%%%%%%%%%%
\section{Exact string solutions in non-constant background fields}
%%%%%%%%%%%%%%%%%%%%%%%%%%%%%%%%%%%%%%%%%%%%%%%%%%%%%%%%%%%%%%%%%%%%%%%%%%%%%
\hspace{1cm} In this section, our aim is to describe several types of exact
solutions for a string moving in background gravitational and NS-NS fields.
To this end, we start with the sigma-model action for $D$-dimensional
space-time
\ba\label{asB} S_1=-\frac{T}{2}\int d^{2}\xi \left[
\sqrt{-\gamma}\gamma^{mn} \p_m X^M\p_n X^N g_{MN}(X) -
\varepsilon^{mn}\p_m X^M\p_n X^N B_{MN}(X)\right],
\\ \nl
\p_m =\p/\p\xi^m,\h \xi^m=(\xi^0,\xi^1)=(\tau,\s),\h m,n=0,1,\h
M,N=0,1,...,D-1, \ea where $T=\left(2\pi\alpha'\right)^{-1}$ is the
(fundamental) string tension, $G_{mn}(X)= \p_m X^M\p_n X^N g_{MN}(X)$,
$B_{mn}(X)=\p_m X^M\p_n X^N B_{MN}(X)$ are the pullbacks of the background
metric and antisymmetric NS-NS tensor to the string worldsheet, and
$\gamma$ is the determinant of the auxiliary metric $\gamma_{mn}$. Varying
(\ref{asB}) with respect to $X^M$ and $\gamma_{mn}$, we obtain the
equations of motion \ba\label{em} &&-g_{LK}\left[\p_m\left(
\sqrt{-\gamma}\gamma^{mn} \p_n X^K\right) +
\sqrt{-\gamma}\gamma^{mn}\Gamma^{K}_{MN}\p_m X^M\p_n X^N \right]
\\ \nl &&= \frac{1}{2}H_{LMN}\varepsilon^{mn}\p_m X^M\p_n X^N, \ea and
the constraints \ba\label{con1}
\left(\gamma^{kl}\gamma^{mn}-2\gamma^{km}\gamma^{ln}\right)\p_m X^M\p_n
X^N g_{MN}\left(X\right)=0,\ea
where  \ba\nl\Gamma^{K}_{MN} =
\frac{1}{2}g^{KL}\left(\p_Mg_{NL} + \p_Ng_{ML} - \p_Lg_{MN}\right)\ea is
the connection compatible with the metric $g_{MN}$ and
\ba\nl H_{LMN}=\p_L B_{MN}+\p_M B_{NL}+\p_N B_{LM}\ea is the $B_{MN}$ field
strength. From now on, we will work in the gauge $\gamma^{mn}=constants$.

First of all, we will try to find a {\it background independent} solution
of the equations of motion of the type
\ba\nl X^M(\xi)=F^M(a_n\xi^n),\h a_n = constants.\ea  It turns out that such
solution exists when $\gamma^{mn}a_m a_n =0$. This leads to
\ba\nl X^M(\xi)=F^M_\pm(u_\pm),\h u_\pm=-\frac{1}{\gamma^{00}}
\left(\gamma^{01}\pm\frac{1}{\sqrt{-\gamma}}\right)\xi^0+\xi^1,\ea
or
\ba\nl X^M(\xi)=F^M_\pm(v_\pm),\h v_\pm=\xi^0+\frac{1}{\gamma^{11}}
\left(-\gamma^{01}\pm\frac{1}{\sqrt{-\gamma}}\right)\xi^1,\ea
where $F^M_\pm$ are arbitrary functions of their arguments. The main
consequence of the obtained result is that for {\it arbitrary background
fields} there exist only one solution, $F^M_+$ or $F^M_-$, but not both at
the same time. In other words, we have only chiral background independent
solutions of the string equations of motion. On the other hand, it must be
noted that these are not solutions of the constraints (\ref{con1}).
Taking a linear combination of the two independent constraints,
we can arrange one of them to be
satisfied, but the other one will give restrictions on the metric. However,
it can be shown that in the zero tension limit, the background independent
solutions of the string equations of motion are also solutions of the
corresponding constraints. Moreover, this result extends to
arbitrary {\it tensionless} $p$-branes \cite{B_PRD60}. It corresponds to
the limit $(-\gamma)^{-1/2}\to 0$, taken in the expressions for $u_\pm$
and $v_\pm$. Let us also note that in conformal gauge, the obtained string
solutions $F^M_\pm(u_\pm)$ and $F^M_\pm(v_\pm)$ reduce to the solutions
$X^M_\pm(\s\pm\tau)$ and $X^M_\pm(\tau\pm\s)$ for left- or right-movers.
These are known to be the only background
independent non-perturbative solutions for an arbitrary static metric,
which are stable and have a conserved topological charge being therefore
topological solitons \cite{MR92}.

A step further is to search for exact solutions of the string equations of
motion {\it and} of the constraints of the type \ba\label{ssv}
X^M(\xi^m)=F^M_\pm(w_\pm) + y(\xi^0),\h \mbox{or}\h
X^M(\xi^m)=F^M_\pm(w_\pm) + z(\xi^1),\ea where $w_\pm=u_\pm$ or
$w_\pm=v_\pm$. The desired property of such ansatz is to allow for the
separation of the variables $\xi^0$ and $\xi^1$. It can easily be shown
that this is achieved only when $F^M_\pm$ are linear functions of $w_\pm$
\ba\nl F^M_\pm(w_\pm)=C^M_\pm w_\pm,\h C^M_\pm = constants.\ea In the zero
tension limit, the condition for separation of the variables $\xi^n$ is
less restrictive and gives \ba\nl F^M(w)=C^M F(w),\h w=u\h\mbox{or}\h
w=v,\ea where $F(w)$ is arbitrary function and we have omitted the
subscripts $\pm$ because in this case $w_+=w_-=w$.

As far as every physically relevant
background metric has some symmetries, let us split the coordinates
$x^M=(x^\mu,x^a)$, $\{\mu\}\neq\{\emptyset\}$ and let us suppose that
there exist an (unspecified) number of independent Killing vectors
$\eta_{\mu}$. Then in appropriate coordinates
$\eta_{\mu}=\p/\p x^{\mu}$ and the metric depends on $x^{a}$ only.
We also assume that the NS-NS 2-form field has at least the same
symmetry, i.e. $\p_{\mu}g_{MN} = \p_{\mu}B_{MN} = 0$. It is natural to
expect that this restrictions will give us the possibility to find
$dim\{\mu\}$ conserved quantities (independent on $\xi^0$ or $\xi^1$
respectively). It turns out this is indeed the case, when
$C^M_\pm=(C^\mu_\pm,0)$. To be more specific, let us use the ansatz
\ba\nl
X^{\mu}\left(\xi^m\right) &=& C^{\mu}_{\pm}u_{\pm} +
y^{\mu}\left(\xi^0\right),
\\ \nl  X^{a}\left(\xi^m\right) &=& y^{a}\left(\xi^0\right).\ea
In order to give a unified description of the tensile and tensionless
string solutions, we parameterize the auxiliary metric $\gamma^{mn}$ as
follows:
\ba\nl \gamma^{00}=-1,\h \gamma^{01}=\lambda^1,\h
\gamma^{11}=(2\lambda^0 T)^2 - (\lambda^1)^2.\ea
Then the Euler-Lagrange equations and the {\it independent} constraints
take the form \cite{B_PRD64} (the overdot is used for $\p/\p\xi^0$)
\ba\nl g_{KL}\ddot{y}^L + \Gamma_{K,MN}\dot{y}^M\dot{y}^N +
2\lambda^0 TC^{\mu}_{\pm}\left(H_{K\mu N} \pm 2\Gamma_{K,\mu
N}\right)\dot{y}^N = 0,\ea
\ba \label{tc1'} g_{MN}(y^a)\dot{y}^M \dot{y}^N = 0,
\\ \label{tc2} C^{\mu}_{\pm} \left[g_{\mu N}(y^a)\dot{y}^N
\pm 2\lambda^0 TC^{\nu}_{\pm} g_{\mu\nu}(y^a)\right] = 0.\ea
The corresponding conserved quantities are
\ba\label{cmB}
g_{\mu N}\dot{y}^N + 2\lambda^0 TC_{\pm}^{\nu}\left(B_{\mu\nu} \pm
g_{\mu\nu}\right) = B^{\pm}_{\mu} = constants.\ea They
are compatible with the constraint (\ref{tc2}) when $B^{\pm}_{\mu}
C^{\mu}_{\pm} = 0$. Using (\ref{cmB}), the equations of motion for $y^a$
and the other constraint (\ref{tc1'}) can be transformed into
\ba\label{emaB} &&2\left(h_{ab}\dot{y}^b\right)^{.} -
\left(\p_a h_{bc}\right)\dot{y}^b\dot{y}^c + \p_a V^{\pm}_B = 4\p_{[a}
B^{\pm}_{c]} \dot{y}^c,\\ \nl &&h_{ab}\dot{y}^a\dot{y}^b +
V^{\pm}_B =0, \ea where
\ba\nl &&h_{ab}\equiv g_{ab} - g_{a\mu}(g^{-1})^{\mu\nu}g_{\nu b},
\\ \nl
&&V^{\pm}_B\equiv \left(2\lambda^0
T\right)^2 C^{\mu}_{\pm}C^{\nu}_{\pm}g_{\mu\nu} + \left(B^{\pm}_{\mu} -
2\lambda^0 TB_{\mu\lambda}C^{\lambda}_{\pm}\right)
(g^{-1})^{\mu\nu}
\left(B^{\pm}_{\nu} - 2\lambda^0 TB_{\nu\rho}C^{\rho}_{\pm}\right),\\ \nl
&&B^{\pm}_a \equiv g_{a\mu}(g^{-1})^{\mu\nu}B^{\pm}_{\nu} + 2\lambda^0
T\left(B_{a\lambda}-g_{a\mu}(g^{-1})^{\mu\nu}B_{\nu\lambda}\right)
C^{\lambda}_{\pm}.\ea
If we restrict the metric $h_{ab}$ to be a diagonal one, i.e.
\ba\nl g_{ab} = g_{a\mu}(g^{-1})^{\mu\nu}g_{\nu b}, \h\h \mbox{for}\h\h
a\ne b,\ea the equations (\ref{emaB}) can be further transformed
into
\ba\nl
\left(\left(h_{aa}\dot{y}^a\right)^2\right)^{.} + \dot{y}^a\p_a\left(h_{aa}
V^{\pm}_B\right) + \dot{y}^a\sum_{b\ne a}
\left[\p_a\left(\frac{h_{aa}}{h_{bb}}\right)\left(h_{bb}\dot{y}^b\right)^2
- 4\p_{[a}B^{\pm}_{b]} h_{aa}\dot{y}^b\right] = 0\ea
(there is no summation over $a$).
Now it is evident that we can always integrate these equations, if all
coordinates on which the background fields depend, except one, are kept fixed.
In the general case, additional restrictions on the metric and on the Kalb-Ramond
field will arise, to ensure the separation of the variables $y^a$ \cite{B_PRD64}.
These conditions are such that the metric and the $B$-field are general enough
to include many interesting cases of (super)string backgrounds in
different dimensions. If the solutions for $y^a$ are already known, the
solutions for $y^\mu$ are obtainable from (\ref{cmB}).
Let us finally note that in the framework of this approach,
exact $D$-string solutions are also found \cite{B_PRD64}.

Our next step is to search for {\it non-chiral} solutions of the string
equations of motion and constraints, i.e. solutions of the type
\ba\label{ncs} X^M(\xi^m)=F^M_+(w_+) + F^M_-(w_-).\ea
In the particular case $B_{MN}=0$, and in conformal and light-cone gauges,
it is known that such solutions do exist \cite{MR92}. To describe them,
we split the string coordinates
\ba\nl X^i=(X^A,X^\alpha),\h A=1,...,D'-1,\h \alpha = D',...,D-2,\ea
and assume that $X^A$ are left-movers, and $X^\alpha$ are right-movers
\ba\label{slrm} X^A\equiv X^A_+,\h X^\alpha\equiv X^\alpha_-,\h
(\p_0\mp\p_1)X_\pm=0.\ea
The metric is supposed to be of the form
\ba\nl g_{MN}=
\left(\begin{array}{cccc}0&1&0&0\\1&0&0&0\\
0&0&g_{AB}&g_{A\beta}\\
0&0&g_{\alpha B}&g_{\alpha\beta}\end{array}\right).\ea
Then, the background allowing both movers is given by
\ba\nl g_{AB}=g_{AB}(X^A_+),\h g_{\alpha\beta}=g_{\alpha\beta}(X^\alpha_-),\h
g_{\alpha B}=\p_\alpha\p_B f(X^A_+,X^\alpha_-),\ea
with $f(X^A_+,X^\alpha_-)$ being an arbitrary function.
Obviously, there are no non-chiral coordinates in this solution: the different
chirality is associated with different string coordinates.

Now, we are going to show that there exist exact solutions of the string
equations of motion and constraints in non-constant background fields
$g_{MN}$ and $B_{MN}$, which posses non-chiral coordinates.
Putting (\ref{ncs}) in the equations of motion (\ref{em}), we obtain the conditions
for the existence of such solutions:
\ba\label{ncsc} \left(2\Gamma_{L,MN}+H_{LMN}\right)\frac{dF^M_+}{dw_+}
\frac{dF^N_-}{dw_-}=0.\ea
For simplicity, we will consider the case when $g_{MN}$ and $B_{MN}$ depend
on only one coordinate, say $r$, and will give the results in conformal gauge.

In our first example, we fix all string coordinates $X^M$ except $X^0$ and $r$
(the remaining coordinates are denoted as $X^\alpha$).
Then the conditions (\ref{ncsc}) and constraints (\ref{con1}) reduce to the
system of equations
\ba\nl &&\p_r g_{00}[(\p_0 X^0)^2 - (\p_1 X^0)^2] - \p_r g_{rr}
[(\p_0 r)^2 - (\p_1 r)^2]=0, \\ \nl
&&\p_r B_{0\alpha}(\p_0 X^0\p_1 r-\p_1 X^0\p_0 r)+\p_r g_{0\alpha}
(\p_0 X^0 \p_0 r-\p_1 X^0 \p_1 r)\\ \nl
&&+\p_r g_{r\alpha}[(\p_0 r)^2-(\p_1 r)^2]=0,\\ \nl
&&\p_r g_{00}(\p_0 X^0\p_0 r-\p_1 X^0\p_1 r)+
\p_r g_{0r}[(\p_0 r)^2 - (\p_1 r)^2]=0, \\ \nl
&&g_{00}[(\p_0 X^0)^2 + (\p_1 X^0)^2] + g_{rr}[(\p_0 r)^2 + (\p_1 r)^2]+
2g_{0r}(\p_0 X^0\p_0 r+\p_1 X^0\p_1 r)=0,\\ \nl
&&g_{00}\p_0 X^0 \p_1 X^0+g_{rr}\p_0 r \p_1 r+
g_{0r}(\p_0 X^0\p_1 r+\p_1 X^0\p_0 r)=0.\ea
Among the nontrivial solutions of the above system, there exist the following
non-chiral ones:
\ba\nl &&\frac{g_{rr}}{g_{00}}=\frac{\p_r g_{rr}}{\p_r g_{00}}=
-\left(\frac{\p_r g_{r\alpha}}{\p_r B_{0\alpha}}\right)^2,\h g_{0r}=0,\\ \nl
&&\p_r B_{0\alpha}\p_0X^0=\p_r g_{r\alpha}\p_1 r,\h
\p_0 X^0 \p_0 r=\p_1 X^0 \p_1 r;\ea
\ba\nl &&\frac{\p_r g_{r\alpha}}{\p_r g_{0\alpha}}=
\frac{\p_r g_{0r}}{\p_r g_{00}},\h
\p_r g_{rr}=\frac{(\p_r g_{0r})^2}{\p_r g_{00}},\h
g_{rr}=\left(2g_{0r}-g_{00}\frac{\p_r g_{0r}}{\p_r g_{00}}\right)
\frac{\p_r g_{0r}}{\p_r g_{00}},\\ \nl
&&\p_r g_{00}\p_0 X^0=-\p_r g_{0r} \p_0 r,\h
\p_0 X^0 \p_1 r=\p_1 X^0 \p_0 r;\ea
\ba\nl &&\frac{\p_r g_{00}}{g_{00}}=
\frac{\p_r g_{0r}}{g_{0r}}=\frac{\p_r g_{rr}}{g_{rr}},\\ \nl
&&-\left[\p_r g_{0\alpha} (\p_0 X^0 \p_0 r
-\p_1 X^0\p_1 r)+\p_r B_{0\alpha}(\p_0 X^0 \p_1 r
-\p_1 X^0\p_0 r)\right]/\p_r g_{r\alpha}\\ \nl
&&=-\p_r g_{00}(\p_0 X^0 \p_0 r -\p_1 X^0\p_1 r)/\p_r g_{0r}\\ \nl
&&=\p_r g_{00}\left[(\p_0 X^0)^2-(\p_1 X^0)^2\right]/\p_r g_{rr}\\ \nl
&&=(\p_0 r)^2-(\p_1 r)^2 .\ea

In our second example, all string coordinates are kept fixed except
$X^0$, $X^1$ and $r$. Only to have readable final expressions, we restrict
the metric to be diagonal, and the NS-NS field to be constant. The solutions
of the corresponding equations following from (\ref{ncsc}) and (\ref{con1})
are
\ba\nl &&\p_0 X^0=+f\p_1 r,\h \p_1 X^0=+f\p_0 r,\h
\p_0 X^1=+h\p_1 r,\h \p_1 X^1=+h\p_0 r;\\ \nl
&&\p_0 X^0=+f\p_1 r,\h \p_1 X^0=+f\p_0 r,\h
\p_0 X^1=-h\p_1 r,\h \p_1 X^1=-h\p_0 r;\\ \nl
&&\p_0 X^0=-f\p_1 r,\h \p_1 X^0=-f\p_0 r,\h
\p_0 X^1=+h\p_1 r,\h \p_1 X^1=+h\p_0 r;\\ \nl
&&\p_0 X^0=-f\p_1 r,\h \p_1 X^0=-f\p_0 r,\h
\p_0 X^1=-h\p_1 r,\h \p_1 X^1=-h\p_0 r,\ea
where
\ba\nl f(g,\p g)=\left(\frac{g_{rr}\p_r g_{11}-g_{11}\p_r g_{rr}}
{g_{11}\p_r g_{00}-g_{00}\p_r g_{11}}\right)^{1/2},\h
h(g,\p g)=\left(\frac{g_{00}\p_r g_{rr}-g_{rr}\p_r g_{00}}
{g_{11}\p_r g_{00}-g_{00}\p_r g_{11}}\right)^{1/2}.\ea
As a consequence, one receives from here that the following equalities are
fulfilled
\ba\nl &&(\p_0\pm\p_1)X^0=f(g,\p g)(\p_0\pm\p_1)r,\h
(\p_0\pm\p_1)X^1=h(g,\p g)(\p_0\pm\p_1)r.\ea
Therefore, we have obtained solutions which allow for all string coordinates
to be non-chiral.

Till now, we considered three types of exact string solutions in variable
external fields $g_{MN}$ and $B_{MN}$. Now, let us see which of them are
compatible with the boundary conditions arising in the open string -
D-brane system.

%%%%%%%%%%%%%%%%%%%%%%%%%%%%%%%%%%%%%%%%%%%%%%%%%%%%%%%%%%%%%%%%%%%%%%%%%%%%%
\section{The open string - D-brane system in non-constant background fields}
%%%%%%%%%%%%%%%%%%%%%%%%%%%%%%%%%%%%%%%%%%%%%%%%%%%%%%%%%%%%%%%%%%%%%%%%%%%%%
\hspace{1cm} The action for an open string ending on a Dp-brane, in the
presence of background gravitational and NS-NS 2-form field, can be written
as
\ba\label{asd} S_2&=&-\frac{T}{2}\int d^{2}\xi \left[
\sqrt{-\gamma}\gamma^{mn} \p_m X^M\p_n X^N g_{MN}(X) -
\varepsilon^{mn}\p_m X^M\p_n X^N B_{MN}(X)\right] \\ \nl
&&-\frac{T}{2}\int d^{2}\xi\varepsilon^{mn}\p_m Y^\mu\p_n Y^\nu F_{\mu\nu}(Y),
\h F_{\mu\nu}=\p_\mu A_\nu - \p_\nu A_\mu,\h (\mu,\nu=0,1,...,p),\ea
where $Y^\mu(\xi)$ are the coordinates on the D-brane, and $A_\mu$ is the
$U(1)$ gauge field living on the D-brane worldvolume. In {\it static gauge}
for the D-brane, one identifies $Y^\mu$ with $X^\mu$. Then the
action (\ref{asd}) acquires the form of the action (\ref{asB}), where
instead of $B_{MN}$ the field $B'_{MN}$ is present. The latter is given by
the equality:
\ba\nl B'_{MN}= B_{MN} - \delta_M^\mu \delta_N^\nu F_{\mu\nu}.\ea
The replacement $B_{MN}\to B'_{MN}$ does not change the equations of motion
(\ref{em}), because $dF=d^2A=0$. The constraints (\ref{con1}) also remain the
same. However, the field $B'_{MN}$ explicitly appears in the expressions for
the generalized momenta
\ba\label{gm} P_M = -T\left(\sqrt{-\gamma}g_{MN}\gamma^{0n}\p_n X^N -
B'_{MN}\p_1 X^N \right),\ea
and in the boundary conditions
\ba\label{mbcs} &&\left[\sqrt{-\gamma}g_{M\nu}\gamma^{1n}\p_n X^\nu
+B'_{M\nu}\p_0 X^\nu\right]_{\s=0,\pi}=0,\\
\label{dbcs} &&X^a(\tau,0)=X^a(\tau,\pi)=q^a,\h a=p+1,...,D-1.\ea
Here we have split the coordinates $X^M$ into $X^\mu$ and $X^a$, and have
denoted the location of the D-brane with $q^a$.

Thus we saw that the exact string solutions, described in the previous
section, are solutions also of the equations of motion and constraints
following from the action (\ref{asd}) in static D-brane gauge. Now we are
going to check their compatibility with the boundary conditions
(\ref{mbcs}), (\ref{dbcs}).

We start by considering the case of background independent solutions
$F^M_\pm(w_\pm)$. On these solutions, the boundary conditions take the
form: \ba\label{bisbcs} \left[\left(g_{M\nu}\pm
B'_{M\nu}\right)\frac{dF^\nu_\pm}{dw_\pm}\right]_{\s=0,\pi}=0,\h
\left[F^a_\pm\right]_{\s=0,\pi}=q^a.\ea Expanding $F^M_\pm$ in Fourier
series one easily checks that they do not give nontrivial solution of the
equations (\ref{bisbcs}). It is clear that the same will be true for the
solutions (\ref{slrm}) with $X^A$left-movers, and $X^\alpha$ right-movers.
As for the solutions of the type (\ref{ssv}), it can be shown that the
conditions (\ref{dbcs}) lead to constant background fields, which is not
the case under consideration. Therefore, to have the possibility to get
nontrivial solutions of the open string - D-brane boundary conditions in
non-constant background fields, we need to have at our disposal exact
string solutions, for which the coordinates are non-chiral. We know from
the previous section that such solutions do exist.

To be able to solve explicitly the boundary conditions, we assume that
$g_{MN}$ and $B'_{MN}$ are constant at $\s=0,\pi$. This is automatically
achieved if $g_{MN}$ and $B_{MN}$ depend only on $X^a$, and the $U(1)$ field
strength $F_{\mu\nu}$ is constant.

From now on we will work in conformal gauge, and we choose to write down
(\ref{ncs}) in the form
\ba\nl X^M(\tau,\s)=X^M_+(\tau+\s) + X^M_-(\tau-\s).\ea
Using the expansions
\ba\nl X^M_\pm(\tau\pm\s) = q^M_\pm + \alpha^M_{0\pm}(\tau\pm\s)+
i\sum_{k\ne 0}\frac{1}{k}\alpha^M_{k\pm}e^{-ik(\tau\pm\s)},\ea
we find the following solution of (\ref{mbcs}) and (\ref{dbcs})
\ba\nl X^\mu(\tau,\s)&=&q^\mu + \left[\delta^\mu_\nu\tau -
\left(g^{-1}B'\right)^{\mu}{}_{\nu}(q^a)\s\right]a^\nu_{0}\\ \nl
&&+\sum_{k\ne 0}\frac{e^{-ik\tau}}{k}
\left[i\delta^\mu_\nu\cos{(k\s)} -
\left(g^{-1}B'\right)^{\mu}{}_{\nu}(q^a)\sin{(k\s)}\right]a^\nu_{k},\\ \nl
X^a(\tau,\s)&=&q^a + \sum_{k\ne 0}\frac{e^{-ik\tau}}{k}b^a_k\sin{(k\s)},\ea
where
\ba\nl \left(g^{-1}B'\right)^{\mu}{}_{\nu} = g^{\mu M}B'_{M\nu},\h
\alpha^M_{k\pm}=\frac{1}{2}\left(a^M_k \pm b^M_k\right).\ea
This result establishes the correspondence with the known solution of the
boundary conditions in the case of constant background fields \cite{CH98}.

It is clear that a crucial role in treating the open string - D-brane system
in variable external fields is played by the conditions (\ref{ncsc}), which
ensure the existence of nontrivial solutions of the type (\ref{ncs}).
Actually, (\ref{ncsc}) are the equations of motion for such type of string
solutions. However, they do not contain second derivatives.
That is why, we propose to consider them as {\it additional constraints}
in the Hamiltonian description of the considered dynamical system.
So, let us compute the resulting constraint algebra.

Using the manifest expression (\ref{gm}) for the momenta, we obtain
the following set of constraints $(\p X\equiv \p X/\p\s)$
\ba\nl &&I_0\equiv g^{MN}P_M P_N-2T\left(g^{-1}B'\right)^{M}{}_{N}P_M \p X^N
+T^2\left(g-B'g^{-1}B'\right)_{MN}\p X^M\p X^N,\\ \nl
&&I_1\equiv P_N\p X^N -
Tg_{MK}\left(g^{-1}B'\right)^{K}{}_{N}\p X^M\p X^N = P_N\p X^N,\\ \nl
&&I_L\equiv \Gamma_{L,MN}g^{MS}g^{NK}P_S P_K
-T\left[2\Gamma_{L,MN}\left(g^{-1}B'\right)^{N}{}_{K}+H_{LMK}\right]
g^{MS}P_S\p X^K\\ \nl
&&+T^2\left\{\Gamma_{L,MN}\left[\left(g^{-1}B'\right)^{M}{}_{S}
\left(g^{-1}B'\right)^{N}{}_{K} - \delta^M_S \delta^N_K\right]
+ H_{LMS}\left(g^{-1}B'\right)^{M}{}_{K}\right\}\p X^S \p X^K .\ea
These constraints have one and the same structure. Namely, all of them are
particular cases of the expression
\ba\nl
I_J \equiv K_J^{SK}(g,\p g)P_S P_K
+ S_{JK}^S(g,\p g,B',\p B')P_S\p X^K
+ R_{JSK}(g,\p g,B',\p B')\p X^S \p X^K,\ea
where $J=(n,L)$, and the coefficient functions $K_J^{SK}$, $ S_{JK}^S$
and $R_{JSK}$ depend on $X^N$ and do not depend on $P_N$. The computation
of the Poisson brackets, assuming canonical ones for the coordinates and
momenta, gives
\ba\label{pba}
\left\{I_{J_1}(\s_1), I_{J_2}(\s_2)\right\} =
\left[M^K_{(J_1}N_{J_2)K}(\s_1)+M^K_{(J_1}N_{J_2)K}(\s_2)\right]
\p\delta(\s_1-\s_2)+C_{[J_1 J_2]}\delta(\s_1-\s_2).\ea
Obviously, the algebra does not close on $I_J$. On the other hand, the
right hand side is quadratic with respect to the newly appeared structures
$M^S_J$ and $N_{JS}$. They are given by
\ba\nl
M^S_J = 2K^{SN}_J P_N + S^S_{JN} \p X^N,\h
N_{JS} = S^M_{JS} P_M + 2R_{JSM}\p X^M,\ea
and satisfy the following Poisson brackets among themselves
\ba\nl
&&\left\{M^{S_1}_{J_1}(\s_1), M^{S_2}_{J_2}(\s_2)\right\}=
\biggl[\left(K^{S_1 N}_{J_1}S^{S_2}_{J_2 N}+
K^{S_2 N}_{J_2}S^{S_1}_{J_1 N}\right)(\s_1)\\ \nl
&&+\left(K^{S_1 N}_{J_1}S^{S_2}_{J_2 N}+
K^{S_2 N}_{J_2}S^{S_1}_{J_1 N}\right)(\s_2)\biggr]
\p\delta(\s_1-\s_2)+C^{S_1 S_2}_{J_1 J_2}\delta(\s_1-\s_2),\\ \nl
&&\left\{N_{J_1 S_1}(\s_1), N_{J_2 S_2}(\s_2)\right\}=
\biggl[\left(S^{N}_{J_1 S_1}R_{J_2 S_2 N}+
S^{N}_{J_2 S_2}R_{J_1 S_1 N}\right)(\s_1)\\ \nl
&&+\left(S^{N}_{J_1 S_1}R_{J_2 S_2 N}+
S^{N}_{J_2 S_2}R_{J_1 S_1 N}\right)(\s_2)\biggr]
\p\delta(\s_1-\s_2)+C_{J_1 J_2 S_1 S_2}\delta(\s_1-\s_2),\\ \nl
&&\left\{M^{S_1}_{J_1}(\s_1), N_{J_2 S_2}(\s_2)\right\}=
\biggl[\left(2K^{S_1 N}_{J_1}R_{J_2 S_2 N}+
\frac{1}{2}S^{S_1}_{J_1 N}S^{N}_{J_2 S_2}\right)(\s_1)\\ \nl
&&+\left(2K^{S_1 N}_{J_1}R_{J_2 S_2 N}+
\frac{1}{2}S^{S_1}_{J_1 N}S^{N}_{J_2 S_2}\right)(\s_2)\biggr]
\p\delta(\s_1-\s_2)+C^{S_1}_{J_1 J_2 S_2}\delta(\s_1-\s_2).\ea
$M^S_J$ and $N_{JS}$ act on $P_M$ and $\p X^M$ as follows
\ba\nl
&&\left\{M^{S}_{J}(\s_1), P_M(\s_2)\right\}= S^S_{JM}(\s_2)
\p\delta(\s_1-\s_2)\\ \nl
&&+\left[2\p_M K^{SN}_J P_N +
\left(\p_M S^S_{JN}-\p_N S^S_{JM}\right)\p X^N\right]\delta(\s_1-\s_2),
\\ \nl
&&\left\{N_{JS}(\s_1), P_M(\s_2)\right\}= 2R_{JSM}(\s_2)
\p\delta(\s_1-\s_2)\\ \nl
&&+\left[\p_M S^N_{JS} P_N +
2\left(\p_M R_{JSN}-\p_N R_{JSM}\right)\p X^N\right]\delta(\s_1-\s_2),
\\ \nl
&&\left\{M^{S}_{J}(\s_1),\p X^M(\s_2)\right\}= 2K^{SM}_{J}(\s_2)
\p\delta(\s_1-\s_2)-2\p_N K^{SM}_J \p X^N\delta(\s_1-\s_2),
\\ \nl
&&\left\{N_{JS}(\s_1),\p X^M(\s_2)\right\}= S^{M}_{JS}(\s_2)
\p\delta(\s_1-\s_2)-\p_N S^{M}_{JS} \p X^N\delta(\s_1-\s_2).\ea
Actually, $I_J$ can be expressed in terms of $M^S_J$ and $N_{JS}$ as
\ba\nl I_J = \frac{1}{2}\left(M^K_J P_K + N_{JK} \p X^K\right).\ea

Let us now see how from the above {\it open} algebra the
{\it closed} algebra of the constraints arises.
For the gauge generators $I_n$, we have
\ba\nl I_0 :
&&M^M_0=2\left[g^{MN}P_N-T\left(g^{-1}B'\right)^{M}{}_{N}\p X^N\right],\\ \nl
&&N_{0M}=2T\left[\left(B'g^{-1}\right)_{M}{}^{N}P_N +
T\left(g-B'g^{-1}B'\right)_{MN}\p X^N\right];\\ \nl
I_1 :
&&M^M_1=\p X^M,\h N_{1M}=P_M.\ea
Inserting these expressions in (\ref{pba}) one obtains
\ba\nl
&&\left\{I_{0}(\s_1), I_{0}(\s_2)\right\} =
(2T)^2\left[I_1(\s_1)+I_1(\s_2)\right]\p\delta(\s_1-\s_2),\\ \nl
&&\left\{I_{1}(\s_1), I_{1}(\s_2)\right\} =
\left[I_1(\s_1)+I_1(\s_2)\right]\p\delta(\s_1-\s_2),\\ \nl
&&\left\{I_{0}(\s_1), I_{1}(\s_2)\right\} =
\left[I_0(\s_1)+I_0(\s_2)\right]\p\delta(\s_1-\s_2).\ea
In this way, we reproduced an old result stating that the string constraint
algebra in a gravitational and 2-form gauge field background coincides with
the one in flat space-time \cite{AO87}.

%%%%%%%%%%%%%%%%%%%%%%%%%%%%%%%%%%%%%%%%%%%%%%%%%%%%%%%%%%%%%%%%%%%%%%%%%%%%%
\section{Concluding remarks}
%%%%%%%%%%%%%%%%%%%%%%%%%%%%%%%%%%%%%%%%%%%%%%%%%%%%%%%%%%%%%%%%%%%%%%%%%%%%%
\hspace{1cm} In this letter we considered the problem of compatibility of
the exact probe string solutions in curved backgrounds with torsion, with
the mixed boundary conditions arising in the open string - Dp-brane
system. We found that there exist solutions of the string equations of
motion and constraints, which also solve these boundary conditions
non-trivially, and give the known result in the constant background fields
limit. We put forward the idea that the conditions for the existence of
such solutions can be considered as a set of additional constraints and
compute their Poisson bracket algebra.

The second part of this work will be devoted to the Hamiltonian analysis of
the open string D-brane system in variable external fields in the framework
of Batalin-Fradkin-Vilkovisky approach, including the issue of the
appearance of a non-canonical Poisson structure at the string endpoints.

\vspace*{.5cm}
{\bf Acknowledgments}
\vspace*{.2cm}

The author would like to acknowledge the hospitality of the ICTP-Trieste,
where this investigation has been done. This work is supported by the
Abdus Salam International Center for Theoretical Physics, Trieste, Italy,
and by a Shoumen University grant under contract $No.20/2001$.

%\newpage
%%%%%%%%%%%%%%%%%%%%%%%%%%%%%%%%%%%%%%%%%%%%%%%%%%%%%%%%%%%%%%%%%%%%%%%%%%%%%%%

\end{document}